\documentclass[11pt,a4paper]{article}
\RequirePackage{amsmath,amssymb}
\RequirePackage[dvipsnames,usenames]{color}

\usepackage{cite}
\usepackage{fullpage}

\usepackage[british]{babel}
\usepackage[latin1]{inputenc}
\usepackage[T1]{fontenc}
\usepackage[final]{showkeys} 
\usepackage[bookmarks]{hyperref}
\usepackage{amsthm}
\usepackage{graphicx}
\usepackage{subfigure}
\usepackage{braket}
\usepackage{mathrsfs}
\usepackage{color}
\usepackage{mathrsfs}
\usepackage{amssymb, bm}

\newcommand{\be}{\begin{equation}}
\newcommand{\ee}{\end{equation}}

\usepackage{bm}
\usepackage{amsfonts}
\usepackage{braket}
\usepackage{graphics}

\newcommand{\bc}{\begin{center}}
\newcommand{\ec}{\end{center}}

\def\theequation{\arabic{section}.\arabic{equation}}

\title{\bf Quasi-geodesics in relativistic gravity}

\author{Valerio~Faraoni and Genevi\`eve Vachon  
$\,$
\\
\\
Department of Physics \& Astronomy, Bishop's University
\\
2600 College Street, Sherbrooke Qu\'ebec, Canada J1M~1Z7
}

\begin{document}
\maketitle

\begin{abstract}

A four-force parallel to the trajectory of a massive particle can always 
be eliminated by going to an affine parametrization, but the affine 
parameter is different from the proper time. The main application is to 
cosmology, in which elements of the cosmic fluid are subject to a pressure 
gradient parallel to their four-velocities. Natural implementations of 
parallel four-forces occur when the particle mass changes and in 
scalar-tensor cosmology.

\end{abstract}

\newpage

\section{Introduction}
\label{sec:1}
\setcounter{equation}{0}

In General Relativity (GR), massive test particles follow timelike 
geodesics and particles subject to (non-gravitational) forces deviate from 
geodesic trajectories. For a freely falling test particle with constant 
mass $m>0$, let $u^a$ be 
the four-tangent to the worldline described by this particle in 
four-dimensional 
spacetime, normalized to $u_a u^a=-1$ (we follow the notations and 
conventions of Ref.~\cite{Wald}). The equation of geodesic curves is
\be
u^b \nabla_b u^a = \alpha(\lambda) u^a \,,\label{non-affinegeodesic}
\ee
where $\lambda $ is a parameter along the trajectory, {\em i.e.},  the 
tangent to a geodesic trajectory is parallely transported. It is always 
possible to change parametrization and use  an affine parameter 
$\sigma$ instead 
(see the Appendix); in this parametrization, the geodesic equation becomes 
\be
u^c \nabla_c u^a = \frac{du^b}{d\sigma}+\Gamma^{a}_{bc} u^b u^c = 
0\,,
\ee
where $\Gamma^a_{bc}$ denotes the Christoffel symbols  and now $u^{\mu} = 
dx^{\mu}/d\sigma$.  The right 
hand side of Eq.~(\ref{non-affinegeodesic}) can always be removed, as one 
would expect on the basis of the Equivalence Principle. In fact, no force 
acts on a geodesic particle and geodesic curves are completely determined 
by the geometry ({\em i.e.}, by gravity), which can be made flat locally 
by going to a freely falling frame; this fact makes one think that all 
appearances of a gravitational force in this frame, including the right 
hand side of Eq.~(\ref{non-affinegeodesic}), are spurious. 

Now, all the other possible affine parameters $\sigma'$ are related to 
$\sigma$ by an affine transformation 
$\sigma \rightarrow \sigma'= a_0 \sigma +b_0 $, where $a_0$ and $b_0$ are 
constants ({\em e.g.}, \cite{Inverno}). In GR, it is customary to use the 
proper time $\tau$ as an affine parameter along timelike geodesics because 
it is the time measured by a clock carried by the observer freely falling 
on that geodesic and is, therefore, a privileged parameter from the 
physical point of view.

A test particle subject to a non-gravitational four-force $F^a$ will 
follow a 
trajectory that deviates from a geodesic and obeys the equation
\be
m \, s^c \nabla_c s^a= F^a 
\ee
(here and in the following, we denote the four-tangent to a geodesic with 
$u^c$ and that to a non-geodesic trajectory with $s^c$). According to 
standard terminology, the particle's four-acceleration $a^b \equiv s^c 
\nabla_c s^b$ is always orthogonal to the four-velocity $s^c$ in the 
4-dimensional 
sense, $a^c s_c=0$, as follows from covariantly differentiating the 
normalization relation $s^c s_c=-1$ and then, if the particle mass $m$ is 
constant,  $F^c =ma^c$ is also 
orthogonal to the particle's trajectory. A force (or a component of the 
acceleration) tangent to the particle's four-velocity $s^c$ can always be 
eliminated by a reparameterization, whether it is of gravitational origin 
({\em i.e.}, pure geometry showing up because of a non-affine 
parametrization) or of non-gravitational nature. The procedure to remove 
the right hand side of Eq.~(\ref{non-affinegeodesic}) (reported in the 
Appendix) does not care 
whether the tangent force is gravitational or not and applies to all 
tangent forces. We have in mind a specific situation occurring in 
cosmology, which will be discussed in Sec.~\ref{sec:2}. If the 
force $F^a$ is non-gravitational, the function $\alpha(\lambda)$ and, 
consequently, the transformation to the geodesic proper time, will depend 
on the particle mass $m$ since the Equivalence Principle is unique to 
gravity and does not apply to non-gravitational forces.

The trajectories of particles subject to forces {\em parallel} 
to the four-tangent to the trajectory are, from the purely mathematical 
point of 
view, geodesic curves and can be affinely parametrized. If one begins with 
the equation
\be
\frac{ds^a}{d\tau_c} +\Gamma^a_{bc} s^b s^c = \alpha(\tau_c) \, s^a \,, 
\ee 
where $\tau_c$ is the proper time of the particle ({\em i.e.}, the time 
measured by a clock at rest with respect to the particle), the affine 
parameter $\tau$ that 
removes the force will be different from $\tau_c$. In other words, the 
proper time $\tau_c$ of the particle subject to a  force and the proper 
time $\tau$ of  
the timelike geodesic obtained from it will differ. Since the four-force 
$F^a $ has no component orthogonal to $s^a$ (in the 4-dimensional 
sense), 
the 
tangent $s^c$ to the 
particle trajectory (parametrized by the proper time $\tau_c$) can 
only deviate from the four-tangent $u^c$ to the corresponding geodesic
 in the time component and in the spatial component parallel to $s^i$ in the 
3-dimensional sense. 
In a local chart $\left\{ x^{\mu} \right\}$, we have
\begin{eqnarray}
s^{\mu} & equiv &\frac{dx^{\mu}}{d\tau_c} = \left( s^0, 
\vec{s} \right)= \frac{dx^{\mu}}{d\tau}\, 
\frac{d\tau}{d\tau_c} \equiv \frac{d\tau}{d\tau_c} \, u^{\mu} \equiv 
\gamma \, u^{\mu} \nonumber\\
&&\nonumber\\
& &=\gamma \left( u^0, \vec{v} \right) \,, 
\end{eqnarray}
where $\gamma (v) \equiv d\tau/d\tau_c$ is  the instantaneous Lorentz 
factor of the Lorentz boost relating the freely falling observer (which 
moves with 3-velocity $\vec{v}$ in the spatial direction of the 
trajectory) and 
the particle subject to the force $F^a$. The 3-spaces perceived by the 
observers $s^a$ and $u^a$ have Riemannian metrics
\begin{eqnarray}
h_{ab} &=& g_{ab} + s_as_b =  g_{ab} + \left( \frac{d\tau}{d\tau_c} 
\right)^2 u_a u_b  \,, \label{hab}\\
&&\nonumber\\
\gamma_{ab} &=& g_{ab} + u_a u_b \,,\label{gammaab}
\end{eqnarray}
respectively. In order to eliminate the 
parallel force from the motion of the particle, one has to abandon its 
proper time as the parameter along the trajectory and adopt the affine 
parameter instead, which is equivalent to a Lorentz boost in the spatial 
direction of the trajectory in 3-space by a Lorentz factor dependent on the 
position along the trajectory. This means shifting time intervals, 
lengths,  
frequencies, and 
energies with respect to the frame comoving with the particle. Therefore, 
although from the mathematical point of view there is no difference 
between a geodesic and the trajectory of a particle subject to a parallel 
force, from the physical point of view there is a fundamental difference 
consisting of the fact that the affine parameter eliminating the force 
is not the proper time which is the physically preferred parameter along 
the trajectory. In other words, the two worldlines correspond to different 
physical observers. We propose to call ``quasi-geodesics'' the timelike 
trajectories of massive particles  subject to a force parallel to the 
trajectory's four-tangent, for which the proper time is not an affine 
parameter.


Several coordinates systems used in GR are based on timelike geodesics ({\em 
i.e.}, on freely-falling observers) and are used in the context of black hole 
physics and horizon thermodynamics. In the Schwarzschild geometry, 
Painlev\'e - Gullstrand coordinates \cite{Painleve,Gullstrand} are based 
on 
radial timelike geodesics. They correspond to the coordinates attached to 
freely falling observers released from rest from infinity and traveling 
radially inward. The more general Martel-Poisson family of coordinate systems 
is obtained when freely falling radial observers are released with non-zero 
initial velocity \cite{MartelPoisson}. This family of cooordinates contains 
Painlev\'e-Gullstrand coordinates as a special case and has as a limit the 
more familiar Eddington-Finkelstein coordinates \cite{Eddington,Finkelstein}. 
Painlev\'e-Gullstrand and Martel-Poisson coordinates have been generalized to 
arbitrary static and asymptotically flat black hole spacetimes in 
\cite{MartelPoisson} and to de Sitter and other static universes in 
\cite{Parikh,FaraoniVachon}. Other coordinates based on radial timelike 
geodesics in the Schwarzschild geometry are the Novikov and the 
Gautreau-Hoffman coordinates, corresponding to observers launched at a 
finite 
radius \cite{MTW,GautreauHoffman}.\footnote{Other familiar coordinates in 
black hole spacetimes are based on radial {\em null} geodesics, including the 
Kruskal-Szekeres coordinates \cite{Kruskal,Szekeres}.}

\section{Perfect fluids} 
\setcounter{equation}{0}
\label{sec:2}

Let us consider a perfect fluid described by the stress-energy tensor
\be
T_{ab}=\left( P+\rho \right) s_a s_b +P g_{ab} \,,\label{perfectfluid}
\ee
where $g_{ab}$ is the metric tensor, $\rho$ is the energy density, $P$ is 
the pressure, and $s^c$ is the fluid four-velocity (normalized to $s_c 
s^c=-1$). The covariant conservation equation $\nabla^b T_{ab}=0$ for 
this perfect fluid reads
\begin{eqnarray}
&& s_a \left[ s^b \nabla_b \left(P+\rho \right) \right] 
+\left(P+\rho\right) 
s^b \nabla_b s_c +\left(P+\rho\right) s_a \nabla^b s_b +\nabla_a P 
\nonumber\\
&&\nonumber\\
&& =0 \,.\label{covcons}
\end{eqnarray}
Observers comoving with the fluid (``comoving observers''), {\em i.e.}, 
with four-velocity $s^a$, perceive the 3-dimensional space as endowed with 
the metric~(\ref{hab}). 
The mixed tensor ${h^a}_b$ is a projector onto this 3-space, because 
$h_{ab} s^a=h_{ab}s^b=0$. Also $\gamma_{ab} s^a=\gamma_{ab} s^b= 
\gamma_{ab} u^a = \gamma_{ab} u^b=0$. By projecting Eq.~(\ref{covcons}) 
along 
the time 
direction $s^a$ of the comoving observers, one obtains the 
time component of the covariant conservation equation
\be
\frac{d\rho}{d\tau_c}  +\left(P+\rho \right) \nabla^b u_b=0  
\,,\label{cons1}
\ee
where $\tau_c$ denotes the proper time of the comoving observers. By 
projecting Eq.~(\ref{covcons}) onto the 3-space ``seen'' by the comoving 
observers ({\em i.e.}, by contracting with the projector ${h^a}_b$), one 
obtains instead 
\be
{h^a}_c \nabla_a P + \left( P+\rho \right) h_{cb} a^b =0 \,,\label{cons2}
\ee
where $a^b \equiv s^c \nabla_c s^b$ is the particle's four-acceleration. 
Let us consider fluid elements, regarded as ``fluid particles''. In 
the absence of external forces, these fluid particles are only subject to 
gravity and to the pressure gradient $\nabla_a P$.  In the case of  dust 
with $P \equiv 0$, the covariant conservation equation $\nabla^b 
T_{ab}=\nabla^b \left(\rho u_a u_b \right)=0$ contains the result that  
dust particles follow geodesics. In fact, the time component of this 
equation gives 
\be
\frac{d\rho}{d\tau} + \rho \nabla^b u_b=0
\ee
which, substituted into the spatially projected conservation equation 
\be
u_a \left( u^b \nabla_b \rho +\rho \nabla^b u_b \right) + \rho u^b \,
\nabla_b u_a=0 
\ee
produces the affinely parametrized geodesic equation $u^b \nabla_b u^a=0$. 
This is the celebrated ``geodesic hypothesis'', {\em i.e.} the result that 
test (or dust) particles follow geodesics. This result is contained in the 
general formalism of GR, does not necessitate a separate assumption 
\cite{Wald}, and contains the additional ingredient that the proper time 
of these test particles is an affine parameter along the geodesics.  

The situation is different for a general perfect fluid with 
pressure. The conservation 
equation~(\ref{cons1}) does not contain the pressure gradient, while the 
spatial equation~(\ref{cons2}) does. If $\nabla_aP$ has a component along 
the time direction (of the comoving observers), {\em i.e.}, along $s^a$, 
this component is annihilated by projecting onto the 3-space orthogonal to 
$s^a$ and Eq.~(\ref{cons2}) will be completely insensitive to it. 
Therefore, the equations of motion of the fluid particles expressed by 
$\nabla^b T_{ab}=0$ are 
completely insensitive to a component of the pressure gradient parallel to 
the four-velocity $s^a$. If the pressure gradient $\nabla^a P$  is exactly  
parallel to $s^a$, it drops out completely from the equations 
describing the trajectories of the fluid particles (the pressure $P$ 
itself still plays a role,  since it gravitates and curves spacetime together 
with the energy 
density $\rho$). Therefore, the particle trajectories may look as if there 
were no forces ({\em i.e.}, geodesic trajectories), but there is  the 
important difference that the proper time $\tau_c$ along the particle 
trajectory does not coincide with the proper time along the corresponding 
geodesic and the observers $s^c$ and $u^c$ perceive different 3-spaces 
Lorentz-boosted with respect to each other.

\section{Cosmology}
\label{sec:3}
\setcounter{equation}{0}

The distinction between freely falling frame and comoving frame along  
a quasi-geodesic trajectory becomes 
essential in cosmology, where the universe is permeated by 
a perfect fluid, or by a mixture of perfect fluids with the same 
four-velocity 
(then the partial densities and the partial pressures simply add up and we 
can consider a stress-energy tensor~(\ref{perfectfluid}) with $\rho$ and $P$ 
equal to the {\em total} energy density and pressure). The spatial 
homogeneity and isotropy of the cosmic microwave 
background and of large scale structures (apart from small perturbations) 
makes the comoving observers assume a privileged role: they are the  
observers who see the cosmic microwave background as 
homogeneous and isotropic around them (apart from tiny temperature 
perturbations $\delta T/T \sim 5 \cdot 10^{-5}$) and cosmological 
observations are usually referred to these observers.

The Friedmann-Lema\^itre-Robertson-Walker (FLRW) line element in comoving 
coordinates $\left(t,r,\vartheta, \varphi \right)$ is 
\be
ds^2 =-dt^2 +a^2(t) \left( \frac{dr^2}{1-kr^2} +r^2 d\Omega_{(2)}^2 \right) 
\,,\label{FLRW}
\ee
where $d\Omega_{(2)}^2 =d\vartheta^2 +\sin^2 \vartheta \, d\varphi^2$ is the line 
element of the unit 2-sphere, $k$ is the curvature index, and the comoving time $t$ 
is the proper time of the comoving observers. We assume that the 
matter source is a perfect fluid described by the stress-energy 
tensor~(\ref{perfectfluid}). The comoving observers coincide with the geodesic 
observers if and only if this fluid is a dust with $P\equiv 0$. 

Let us examine the geodesic equation in the geometry~(\ref{FLRW}). The only 
non-vanishing Christoffel symbols are 
\begin{eqnarray}
\Gamma^0_{11}&=& \frac{a\dot{a}}{1-kr^2} \,, \;\;\;\;\;\; 
\Gamma^0_{22}=a\dot{a}r^2 \,, \;\;\;\;\;\; 
\Gamma^0_{33}=a\dot{a}r^2 \sin^2 \vartheta \,, \\
&&\nonumber\\
\Gamma^1_{01}&=& \Gamma^2_{02}= \Gamma^3_{03} = \frac{\dot{a}}{a}\,, \;\;\;\;\;\;
\Gamma^1_{22}= -r \left( 1-kr^2 \right) \,,\\
&&\nonumber\\
\Gamma^1_{33}&=& -r \left( 1-kr^2\right) \sin^2 \vartheta \,, \;\;\;\;\;\; 
\Gamma^2_{33} =  -\sin\vartheta \cos\vartheta \,, \\
&&\nonumber\\
\Gamma^3_{22}&=& \cot\vartheta \,, \;\;\;\;\;\; 
\Gamma^2_{12}= \Gamma^3_{13}=   \frac{1}{r} \,,
\end{eqnarray}
and those related to them by the symmetry $\Gamma^a_{bc}=\Gamma^a_{cb}$ 
and an overdot denotes 
differentiation with respect to the comoving time $t$.   
Radial timelike geodesics are parametrized by the proper time $\tau$ of 
the freely falling observers and  have 
$u^{\vartheta}=d\vartheta/d\tau= u^{\varphi}=d\varphi/d\tau=0$. 
The radial component of the geodesic equation is 
\be
\frac{du^r}{d\tau} + 2H u^t u^r =0 \,,
\ee
where $H \equiv \dot{a}/a$ is the Hubble function. This equation is  
immediately 
integrated to 
\be
u^r=u^r_{(0)} \, \frac{a_{(0)}^2}{a^2} \,,
\ee
where $u_{(0)}^r \equiv u^r( \tau_0)$ is the initial condition at a point 
$\tau_0$ along the geodesic trajectory, where $a$ assumes the value $a_0$. 
The normalization 
of the 
four-velocity $g_{ab}u^a u^b=-1$ gives its time component as 
\be
u^t =\sqrt{ 1+ \frac{\left( u^r_{(0)} \right)^2 a_{(0)}^4 }{a^2 \left(1-kr^2\right) } 
} \,.
\ee
Let the massive particle (or geodesic observer) be initially at rest at 
the point of comoving coordinates $\left( t_0, r_0, \vartheta, \varphi 
\right)$, or
\be
u^{\mu}_{(0)} =\left( 1, 0,0,0 \right) \,.
\ee
Then, the timelike radial geodesic with this initial condition has tangent 
$u^{\mu} =\left( 1, 0,0,0 \right) $ at all subsequent times. In 
particular, $u^t\equiv dt/d\tau=1$ and comoving and proper time coincide 
(apart from a possible shift in the origin). This is not true for any 
radial timelike geodesics, but only for those satisfying the special 
initial condition $u^a_{(0)}= s^a_{(0)}$. In other words, particles can 
move outward radially at any speed describing the same spacetime 
trajectories, but if their radial velocities are synchronized initially 
with the Hubble flow, they remain synchronized and their proper time then 
coincides with the comoving time.

We can now find the relation between the proper times $t$ and $\tau$ of 
comoving and freely falling ({\em i.e.}, geodesic) observers. The time 
component of the geodesic equation is 
\be
\frac{du^t}{d\tau} + \frac{a_{(0)}^4 \dot{a}}{a^3\left( 1-kr^2\right) }  
\left( u_{(0)}^r \right)^2 =0 
\ee
and the initial condition $u_{(0)}^r=0$ yields 
\be
t(\tau) = C_1\tau + C_2 \,,
\ee
where the $C_{i}$ are integration constants. If the special radial timelike 
geodesics satisfy the initial condition $u^a_{(0)}= s^a_{(0)}$, it is also 
$C_1=1$ and not only the set of spacetime points lying along the geodesic 
curve and the comoving observer's worldline coincide, but also their 
parametrizations coincide.  Thus, if freely falling observers are given a 
special initial velocity that synchronizes them with the Hubble flow 
initially, they remain in the Hubble flow at all subsequent times. Comoving 
time and the proper time of freely falling observers then coincide, but it 
is important to realize that this does not happen for all radial timelike 
geodesics, only for those that satisfy the special (synchronizing) 
initial condition.

\section{Physical nature of a parallel force}
\label{sec:4}
\setcounter{equation}{0}

\subsection{A parallel force cannot be electromagnetic}

A four-force parallel to the particle trajectory cannot be an 
electromagnetic 
force. In fact, the electromagnetic four-force on a test charge $q$ is 
\be
F^{a} = qF^{ab}s_{b}, \,,
\ee
where $s^a$ is the four-tangent to the particle 
worldline and $F_{ab}$ is the Maxwell tensor. In the frame of this 
particle, the components of the four-tangent are $s^{\mu}=\left( 
s^0,0,0,0\right)$ and $F^{\mu}=q F^{\mu 0}s_0$ is purely spatial because, for 
$\mu=i=1,2,3$,  $F_{0i}=- E_i$, where $E^i$ is 
the electric field perceived by the particle. For $\mu=0$, one has $F^{00}=0$ 
due to the antisymmetry of the Maxwell tensor, and the electric force cannot 
have  a time component and must be purely spatial. Similarly, the magnetic 
force cannot have  a time component because the purely spatial magnetic field 
$B^{\mu}$ is built 
out of the space-space component $F_{ij}$ of $F_{\mu\nu}$  
according to
\be
F_{\mu\nu}=\left( \begin{array}{cccc}
0 & -E_x & -E_y & -E_z \\
E_x & 0 & B_z & -B_y \\
E_y & -B_z & 0 & B_x\\
E_z & -B_y & B_z & 0 \\
\end{array} \right)
\ee 
in local Cartesian coordinates (this is even more intuitive, 
since the Lorenz force due to a purely magnetic field is perpendicular to 
the particle's 3-velocity in the 3D sense). Therefore, a force parallel to 
the  worldline of a massive particle cannot be of electromagnetic nature.

\subsection{Parallel force due to a variable particle mass}

A four-force parallel to the trajectory (with four-tangent $s^a$) is akin 
to 
a 
variable particle mass. In fact, the 4-dimensional analogue of Newton's 
second law
\be
F^a =\frac{Dp^a}{D\tau } = \frac{ D(m s^a)}{D\tau } = \frac{dm}{d\tau } 
\, s^a + m \, \frac{Ds^a}{D\tau } \,,
\ee
where $\tau$ is the proper time along the trajectory, can be rewritten as 
\be
\frac{Ds^a}{D\tau } \equiv  \frac{ds^a}{d\tau } 
+\Gamma^a_{bc}s^b s^c  =-\left[ \frac{d }{d\tau } \, \ln \left( \frac{m 
}{m_0} \right)\right]  s^a \,,
\ee
where $m_0$ is a constant with the dimensions of a mass. 
The variation of the mass $m(\tau)$ along the trajectory generates a force 
$F^a = - \, \frac{dm}{d\tau} \, s^a $ parallel to it. This force can be 
eliminated by a reparametrization (as done in the Appendix), however the 
affine parameter $\lambda$ that achieves this is different from the proper 
time $\tau$ measured by the observer.  {\em Vice-versa}, a force parallel 
to the trajectory of a particle can be interpreted as a variation of the 
particle's mass along the trajectory by setting 
\be
\frac{d}{d\tau} \ln \left( \frac{m}{m_0} \right) =- \alpha (\tau) \,, 
\ee
which gives the mass dependence
\be
m(\tau) = m_0 \, \mbox{e}^{-\int d\tau \, \alpha(\tau)} \,.
\ee
By using Eq.~(\ref{A7}), the affine parameter is found to be 
\be
\sigma(\tau) =A \int d\tau \, \frac{m(\tau)}{m_0}  +B \,.
\ee
This discussion applies regardless of the physical process causing the 
variation of the particle mass. Variable masses are encountered, for 
example, in rockets\footnote{Analytical solutions of the Einstein 
equations describing photon rockets have a long history 
\cite{Kinnersley,Bonnor1,Damour,Bonnor2,DainMoreschiGleiser,Cornish1,Cornish2,Podolsky1,Podolsky2}.}  
and in solar sails ({\em e.g.}, \cite{Forward,Fuzfa1,Fuzfa2}).

The variation of particle masses in cosmology 
has been studied in 
Refs.~\cite{DamourGibbonsGundlach90,CasasBellidoQuiros92,Bellido93,AndersonCarroll97,Mbelek98,Mbelek04}, mostly 
in the context of scalar-tensor gravity, which is discussed next.

\subsection{Parallel forces in scalar-tensor cosmology}

The (Jordan frame) Brans-Dicke action is 
\begin{eqnarray}   
S^{(BD)} &=&\frac{1}{16\pi} \, \int d^4 x \, \sqrt{-g} \left[
\phi R -\frac{\omega}{\phi} \, g^{cd} \nabla_c\phi \, \nabla_d\phi
-V( \phi) \right] \nonumber\\
&&\nonumber\\
&\, & + S^{(m)} \,, \label{1}
\end{eqnarray}
where 
\be
S^{(m)}=\int d^4 x \, \sqrt{-g} \, {\cal L}^{(m)}
\ee
is the matter action, $\phi>0$ is the Brans-Dicke scalar field, and the 
dimensionless parameter $\omega $ is the Brans-Dicke coupling. The 
conformal transformation of the metric 
\be
g_{ab}\rightarrow \tilde{g}_{ab}=\Omega^2 \, g_{ab} \,,  \;\;\;\;\;\;\ 
\Omega=\sqrt{\phi} \,,
\ee
and the scalar field redefinition 
\be \label{46} 
\tilde{\phi}( \phi)= \sqrt{ \frac{2\omega+3}{16\pi G} } \,
\ln \left( \frac{\phi}{\phi_0} \right) 
\ee 
(where $\omega > -3/2$) bring the Brans-Dicke action~(\ref{1}) into its 
Einstein frame form \cite{Dicke62} 
\begin{eqnarray} 
S& =& \int d^4 x \, \left\{ \sqrt{ -\tilde{g}} \left[
\frac{ \tilde{R}}{16\pi G} -\frac{1}{2} \, \tilde{g}^{ab}
\tilde{\nabla}_a\tilde{\phi} \tilde{\nabla}_b\tilde{\phi} -U\left(
\tilde{\phi} \right) \right] \right.
\nonumber \\ 
& & \left. + \mbox{e}^{ -8\sqrt{ \frac{\pi G}{2\omega +3} } \,\,
\tilde{\phi} } {\cal L}^{(m)} 
\left[ \tilde{g} \right] \right\} \,,
\label{47} 
\end{eqnarray} 
where $\tilde{\nabla}_a$ is the covariant derivative
operator of the rescaled metric $\tilde{g}_{ab}$,
\be \label{47bis}
U\left( \tilde{\phi} \right) = V\left[ \phi \left( \tilde{\phi} \right)
\right] \exp \left( -8 \sqrt{\frac{\pi G}{2\omega+3} } \, \tilde{\phi}
\right) \,
\ee
and a tilde denotes Einstein frame quantities. 
(The redefinition~(\ref{46}) has the purpose of casting the scalar 
field kinetic energy density into canonical form.)  In the Einstein frame, 
the matter Lagrangian density is  multiplied 
by an  exponential factor with argument proportional to  $\tilde{\phi}$ 
(eq.~(\ref{47})): this scalar couples explicitly to matter. 

In general, the covariant conservation equation for the matter stress-energy 
tensor $\nabla^{b} \, T_{ab}^{(m)} =0 $ 
is not conformally invariant \cite{Wald}: the conformally transformed 
$\tilde{T}_{ab}^{(m)} $ satisfies instead 
\be \label{43}
\tilde{\nabla}^{b} \, \tilde{T}_{ab}^{(m)} =-\,\frac{d}{d\phi}
\left[  \ln \Omega (\phi) \right] \, \tilde{T}^{(m)} \, \tilde{\nabla}_a
\phi \,.
\ee
Only conformally
invariant matter with  $T^{(m)} =0 $ is conformally invariant. 

As done in Sec.~\ref{sec:1} for dust, the Einstein frame modification of the 
geodesic equation can be derived from Eq.~(\ref{43}). Timelike geodesics of 
the original metric $g_{ab}$ are not mapped into geodesics 
of $\tilde{g}_{ab}$ because of the force
proportional to $ \tilde{\nabla}^a \phi$ introduced by the conformal 
transformation.  When applied to the Einstein frame action, the 
stress-energy tensor expression
\be \label{43bis}
\tilde{T}_{ab}^{(m)}=\frac{-2}{\sqrt{ -\tilde{g} } } \, \frac{  \delta
\left( \sqrt{-\tilde{g}} \, \, {\cal L}^{(m)} \right) }{\delta
\tilde{g}^{ab} } \,,
\ee
yields
\begin{eqnarray} 
\tilde{T}_{ab}^{(m)} &=& \Omega^{-2} \, T_{ab}^{(m)} \,, \;\;\;\;
 \widetilde{  { {T_{a}}^{b}}^{(m)} } = \Omega^{-4} \, {{T_a}^b}^{(m)} 
\,,\\
&&\nonumber\\
{\tilde{T}}^{ab} &=& \Omega^{-6} \, {T^{ab}}^{(m)} \,, \label{43ter}
\end{eqnarray}
and
\be \label{43quater}
\tilde{T}^{(m)}= \Omega^{-4} \, T^{(m)} \,.
\ee

For a perfect fluid with stress-energy tensor~(\ref{perfectfluid}), the 
conformal map generates 
\be \label{4305}
\tilde{T}_{ab}^{(m)}=\left( \tilde{P}^{(m)}+\tilde{\rho}^{(m)}
 \right) \tilde{u}_a
\tilde{u}_b +\tilde{P}^{(m)} \, \tilde{g}_{ab} \,,
\ee
in the rescaled world, where $ \tilde{g}_{ab} \, \tilde{u}^a \tilde{u}^b=-1$, 
from which one obtains 
\be \label{4307}
\tilde{u}^a=\Omega^{-1} \, u^a \,, \;\;\;\;\;\;\;
\tilde{u}_a=\Omega \, u_a \,.
\ee
Equations~(\ref{43ter}), (\ref{4305}), and~(\ref{4307}) yield
\begin{eqnarray} 
\left( \tilde{P}^{(m)} + \tilde{\rho}^{(m)} \right) \tilde{u}_a 
 \tilde{u}_b    + \tilde{P}^{(m)} \, 
\tilde{g}_{ab} &=& \Omega^{-2}
\left[ \left( P^{(m)} +\rho^{(m)} \right) u_a u_b \right.\nonumber\\
&&\nonumber\\
& \, &  \left.  + P^{(m)} \, g_{ab} \right]    \,, \label{4308} 
\end{eqnarray}
from which the transformation properties 
\be \label{4309}
\tilde{\rho}^{(m)}=\Omega^{-4} \, \rho^{(m)} \,, \;\;\;\;\;\;\;
\tilde{P}^{(m)}=\Omega^{-4} \, P^{(m)} 
\ee
follow. 
If the Jordan frame fluid is described by the equation of
state 
\be \label{minchia}
P^{(m)}=w \rho^{(m)}
\ee
with $w=$constant, the same equation of state is valid in the Einstein 
frame due to Eq.~(\ref{4309}).

In FLRW spacetimes, the Jordan frame  fluid conservation equation 
\be 
\frac{d \rho^{(m)}}{dt} +3H \left( P^{(m)} +\rho^{(m)} \right)=0
\ee
is mapped into the Einstein frame equation
\begin{eqnarray} 
\frac{d \tilde{\rho}^{(m)}}{dt}+3 \tilde{H} \left( \tilde{P}^{(m)}
+\tilde{\rho}^{(m)}
\right)&=&\frac{d \left( \ln \Omega \right)}{d\phi} \,\, \dot{\phi} \left( 
3\tilde{P}^{(m)} - \tilde{\rho}^{(m)} \right) \,.\nonumber\\
&&
\end{eqnarray}

Let us return to general spacetimes. Under the conformal 
rescaling, the stress-energy tensor
$T_{ab}^{(m)}$ scales according to 
\be 
\tilde{T}^{ab}_{(m)}=\Omega^s \, \, T^{ab}_{(m)} \,, \;\;\;\;\;\;\;\;\;
\tilde{T}_{ab}^{(m)}=\Omega^{s+4} \,\, T_{ab}^{(m)} \,, 
\ee
where $s$ is an appropriate conformal weight and  the Jordan frame covariant 
conservation equation $\nabla^b \,  T_{ab}^{(m)}=0$ maps (in four spacetime 
dimensions) to \cite{Wald,book1}
\begin{eqnarray} 
\tilde{\nabla}_a \left( \Omega^s \, T^{ab}_{(m)} \right) &=&\Omega^s
\, \nabla_a T^{ab}_{(m)} +\left( s+6 \right) \Omega^{s-1} \, 
T^{ab}_{(m)}\nabla_a  \Omega  \nonumber\\ 
&&\nonumber\\
&\, & - \Omega^{s-1} g^{ab} \, T^{(m)} \nabla_a \Omega \,. \label{x1}
\end{eqnarray}
Choosing the conformal weight $s=-6$ yields,
consistently with eq.~(\ref{43quater}),
\be \label{x3}
\tilde{T}^{(m)} \equiv    \tilde{g}^{ab} \, 
\tilde{T}_{ab}^{(m)}= \Omega^{-4}  \,\, T^{(m)} 
\ee
and Eq.~(\ref{x1}) is mapped to 
\be \label{x3bis}
\tilde{\nabla}_a  \tilde{T}^{ab}_{(m)}  =- \tilde{T}^{(m)}\,
\tilde{g}^{ab} \, \tilde{\nabla}_a
\left(  \ln \Omega \right)  \, . 
\ee
For  Brans-Dicke theory with $\Omega=\sqrt{G\phi}$, 
\cite{Wagoner70,book1} 
\be \label{x4}
\tilde{\nabla}_a  \tilde{T}^{ab}_{(m)}  =- \frac{1}{2\phi} \,\,
\tilde{T}^{(m)}
 \,  \tilde{\nabla}^b \phi  = - \sqrt{ \frac{4\pi G}{2\omega
+3} } \, \, \tilde{T}^{(m)} \,\, \tilde{\nabla}^b \tilde{\phi} \,,
\ee
from which one derives the corrected geodesic equation. For a  dust fluid 
with $P=0$, one obtains
\begin{eqnarray} 
&& \tilde{u}_a \, \tilde{u}_b \, \tilde{\nabla}^b  \tilde{\rho}^{(m)} 
+\tilde{\rho}^{(m)} \, \tilde{u}_a \, \tilde{\nabla}^b  \tilde{u}_b
+\tilde{\rho}^{(m)} \, \tilde{u}_c \,\tilde{\nabla}^c \, \tilde{u}_a 
\nonumber\\
&&\nonumber\\
&\, &  - \sqrt{ \frac{4\pi G}{2\omega +3} } \, \,
\tilde{\rho}^{(m)} \, \tilde{\nabla}_a \tilde{\phi} =0 \,. \label{x7}
\end{eqnarray}
In terms of the proper time $\tau$ along the fluid worldlines
with tangent $\tilde{u}^a$, Eq.~(\ref{x7}) reads
\begin{eqnarray}
&& \tilde{u}_a \left(
\frac{d\tilde{\rho}^{(m)} }{d\tau}+\tilde{\rho}^{(m)} \, \tilde{\nabla}^c
\tilde{u}_c \right) +\tilde{\rho}^{(m)} 
\left(  \frac{ d\tilde{u}_a}{d\tau}
-\, \sqrt{ \frac{4\pi G}{2\omega+3}} \, \tilde{\nabla}_a\phi \right) 
\nonumber\\
&&\nonumber\\
&& =0 \,,   \label{x8}
\end{eqnarray}
equivalent to
\be \label{x9}
\frac{d\tilde{\rho}^{(m)} }{d\tau} + \tilde{\rho}^{(m)} \, \tilde{\nabla}^c
\tilde{u}_c =0 
\ee
and
\be \label{x10}
\frac{D\tilde{u}^a }{D\tau} =\sqrt{ \frac{4\pi G}{2\omega+3}} \, \,   
\tilde{\nabla}^a \tilde{\phi} \,.
\ee
The Einstein frame cousin of the geodesic equation is\footnote{A 
similar 
correction to the geodesic equation appears in dilaton gravity 
that results from the low-energy limit of 
string theories, but there the coupling of the dilaton may not be 
universal   
\cite{TaylorVeneziano88,DamourPolyakov94,Gasperini99}.}  
\cite{Wagoner70,Cho92,Cho97}
\be
\frac{d^2 x^a}{d\tau^2} +\tilde{\Gamma}_{bc}^a \,  
\frac{d x^b}{d\tau} \, 
\frac{d x^c}{d\tau} = 
\sqrt{ \frac{4\pi G}{2\omega + 3}} \, \, \tilde{\nabla}^a \tilde{\phi} 
\,.\label{x11} 
\ee

In general spacetimes, $\nabla^c \phi$ does not point along the particle 
trajectory. However, in an unperturbed FLRW universe, $\phi=\phi(t)$ and 
$\nabla^c \phi$ does point in the time direction of comoving observers. In 
this case, we have a force parallel to the worldline of a massive particle 
in the quasigeodesic equation
\be
\frac{d^2 x^a}{d\tau^2} +\tilde{\Gamma}_{bc}^a \,  
\frac{d x^b}{d\tau} \,  \frac{d x^c}{d\tau} = 
\sqrt{ \frac{4\pi G}{2\omega + 3}} \, \frac{d\tilde{\phi}}{d\tau_c}\, 
\frac{d\tau_c}{d\tau} \,  s^a \,,
\ee
where we have identified $\tilde{u}^a$ with $s^a$ to keep with our notation 
of the previous sections. This parallel force in scalar-tensor FLRW cosmology 
can be 
seen as a variation of the particle mass along its trajectory.

\section{Summary and conclusions}
\label{sec:5}
\setcounter{equation}{0}

A geodesic curve can be parametrized affinely or non-affinely. The 
general mathematical definition of a curve is that a curve is an 
equivalence class, where the equivalence relation is a change of 
parametrization, therefore 
affinely- and non-affinely-parametrized geodesics coincide. This 
definition does not take into account the fact that the proper time is a 
physically preferred parameter along the trajectory. The worldline 
of a particle subject to a parallel force may coincide, point by point, 
with a timelike geodesic however, in general, it cannot be affinely 
parametrized  keeping the proper time as the parameter. A 
freely falling frame along the curve does not, in general, 
coincide with the comoving frame associated with the observer subject to a 
parallel force and following the same worldline; the time coordinates of 
these observers,  {\em i.e.}, their proper times $\tau$ and $\tau_c$, do 
not coincide. 

FLRW cosmology is an exception: by setting a special initial condition 
({\em i.e.}, synchronizing the velocity along the quasi-geodesic with the 
Hubble flow), the proper time can be made to coincide with the 
affine parameter. 


One can speculate on particular physical realizations of a force parallel 
to the timelike trajectory of a particle subject to it. A natural 
realization occurs when the particle mass changes along the trajectory 
(for example in rockets and light sails). Another realization occurs in 
scalar-tensor cosmology and string cosmology, where the dilaton field acts 
in a way similar to the Brans-Dicke gravitational scalar field. At 
present, it is not clear whether there are other physically meaningful 
ways of achieving such parallel forces.\\\\

{\small 
\noindent {\bf Acknowledgements.} This work is supported, in part, by the 
Natural Sciences \& Engineering Research Council of Canada (Grant no. 
2016-03803 to V.F.) and by Bishop' s University.  }

\appendix
\section*{Appendix}
\label{appendix}
\renewcommand{\theequation}{A.\arabic{equation}}
\setcounter{equation}{0}

Begin from the non-affinely parametrized geodesic equation
\be
\frac{d^2 x^{\mu}}{d\lambda^2} +\Gamma^{\mu}_{\alpha\beta} \, 
\frac{d x^{\alpha}}{d\lambda}\, \frac{d x^{\beta}}{d\lambda}= 
\alpha(\lambda) \, \frac{dx^{\mu}}{d\lambda} \,.\label{A1}
\ee
A reparametrization  $\lambda \rightarrow \sigma (\lambda) $ produces 
\begin{eqnarray}
\frac{d x^{\mu}}{d\lambda} &=& \frac{d x^{\mu}}{d\sigma} \, \frac{d  
\sigma}{d\lambda} \,,\\
&&\nonumber\\
\frac{d^2 x^{\mu}}{d\lambda^2} &=& 
\left( \frac{d\sigma}{d\lambda} \right)^2  \frac{d^2 x^{\mu}}{d\sigma^2} 
+ \frac{d^2 \sigma}{d\lambda^2}\, 
\frac{dx^{\mu}}{d\sigma} 
\end{eqnarray}
and changes Eq.~(\ref{A1}) into 
\begin{eqnarray}
\left( \frac{d\sigma}{d\lambda} \right)^2 \frac{d^2 x^{\mu}}{d\sigma^2}
+ \frac{d^2 \sigma}{d\lambda^2}\, \frac{dx^{\mu}}{d\sigma} 
+  \left( \frac{d\sigma}{d\lambda} \right)^2 
\Gamma^{\mu}_{\alpha\beta} \, \frac{dx^{\alpha}}{d\sigma} \, 
\frac{dx^{\beta}}{d\sigma} \nonumber\\
\nonumber\\
= \alpha(\lambda) \, 
\frac{d\sigma}{d\lambda} 
\, \frac{dx^{\mu}}{d\sigma} \,.
\end{eqnarray}
The affine parametrization is obtained by imposing that
\be
\frac{d^2\sigma}{d\lambda^2} \, \frac{dx^{\mu}}{d\lambda} =\alpha(\lambda) 
\, \frac{d\sigma}{d\lambda} 
\,  \frac{dx^{\mu}}{d\lambda}  \,,
\ee
which leads to the second order ordinary differential equation
\be
\frac{d^2\sigma}{d\lambda^2}  -\alpha(\lambda)  \frac{d\sigma}{d\lambda}  
=0 
\ee
for the unknown function $\sigma(\lambda)$. This equation always admits the  
solution 
\be
\sigma(\lambda) = A \int d \lambda \, \mbox{e}^{ -\int d\lambda 
\, \alpha(\lambda)} +B \,, \label{A7}
\ee
where $A$ and $B$ are integration constants.

\section*{Acknowledgments}

This work is supported, in part, by the Natural Sciences \& Engineering 
Research Council of Canada (Grant No. 2016-03803) and by Bishop's 
University.

\end{document}